# PCAPVision: PCAP-Based High-Velocity and Large-Volume Network Failure Detection

Lukasz Tulczyjew
lukasz.tulczyjew@b-yond.com
B-Yond
Warsaw, Poland

Ihor Biruk
biruk1230@gmail.com
B-Yond
Wrocław, Poland

Murat Bilgic
Murat.Bilgic@b-yond.com
B-Yond
Redmond, Washington, USA

Charles Abondo
charles.abondo@b-yond.com
B-Yond
Montreal, Quebec, Canada

Nathanael Weill*
nathanael.weill@b-yond.com
B-Yond
Montreal, Quebec, Canada

*corresponding author

## Abstract

Detecting failures via analysis of Packet Capture (PCAP) files is crucial for maintaining network reliability and performance, especially in large-scale telecommunications networks. Traditional methods, relying on manual inspection and rule-based systems, are often too slow and labor-intensive to meet the demands of modern networks. In this paper, we present PCAPVision, a novel approach that utilizes computer vision and Convolutional Neural Networks (CNNs) to detect failures in PCAP files. By converting PCAP data into images, our method leverages the robust pattern recognition capabilities of CNNs to analyze network traffic efficiently. This transformation process involves encoding packet data into structured images, enabling rapid and accurate failure detection. Additionally, we incorporate a continual learning framework, leveraging automated annotation for the feedback loop, to adapt the model dynamically and ensure sustained performance over time. Our approach significantly reduces the time required for failure detection. The initial training phase uses a Voice Over LTE (VoLTE) dataset, demonstrating the model's effectiveness and generalizability when using transfer learning on Mobility Management services. This work highlights the potential of integrating computer vision techniques in network analysis, offering a scalable and efficient solution for real-time network failure detection.

## 1 Introduction

Detecting failures in PCAP files is a critical task in network troubleshooting and analysis. PCAP files record detailed packet-level data, but as networks grow, the volume of data becomes overwhelming. Traditional methods for detection rely heavily on manual inspection and rule-based systems, which are time-consuming and impractical for large-scale networks where timely failure detection is essential. Failures, unlike errors, indicate unsuccessful communication attempts, such as failed calls, which can significantly impact network performance and user experience and require more in-depth analysis. The primary goal of PCAPVision is to quickly identify failed call-flows in PCAP files, allowing further analysis of only relevant data and avoiding the need to store and analyze large volumes of data.

Efficient failure detection in PCAP files is crucial for maintaining network reliability and performance. Minimizing the Mean Time to Detect (MTTD) and Mean Time to Repair (MTTR) ensures seamless service delivery and customer satisfaction. Traditional parsing methods are accurate, but often too slow for real-time network monitoring and troubleshooting. Rapid filtering and analysis of PCAP files can significantly reduce downtime and operational costs. Proactive network management enabled by quick failure detection helps prevent minor issues from becoming major problems, especially in complex modern data centers and distributed systems.

Advanced techniques, such as using Convolutional Neural Networks (CNNs) to analyze transposed images of PCAP files, offer a promising solution. This approach enhances speed and accuracy in detecting failures, identifying subtle failure patterns that traditional methods might miss. Our approach leverages computer vision and CNNs to detect failures by transforming PCAP files into images (See Figure 1) using CNN's pattern recognition capabilities instead of traditional parsing techniques (f.i. using WireShark). This transformation involves mapping packet data into structured images, which the CNN can then process to detect failures.

Computer vision and CNNs excel at pattern recognition, making them ideal for analyzing PCAP files. This approach bypasses the limitations of traditional parsing methods, allowing for rapid filtering and efficient analysis of large volumes of PCAP data. It enables quicker failure identification, enhancing network management efficiency.

In a rapidly changing communications network, a static model has the risk of becoming obsolete very quickly. To address this, our methodology integrates a continual learning framework to maintain robust performance over time (see Figure 3). This framework combats catastrophic forgetting through experience replay and adaptive regularization. Experience replay helps the model retain old tasks by reusing stored data, while adaptive regularization ensures that new information reinforces rather than overwrites existing knowledge. Continual learning is crucial for sustaining high performance amidst rapid technological changes and evolving network traffic patterns.

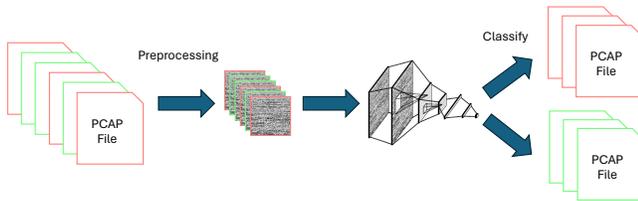

Figure 1: Overview of the process from PCAP file to image conversion and CNN prediction.

This article introduces our approach using computer vision to detect failure in PCAP files and continual learning to maintain performance over time. The key contributions of our research are:

- **Efficient PCAP Processing**: Enables high-velocity failure detections across large volumes of PCAP files, significantly reducing storage and network probe costs.
- **Advanced Pattern Extraction**: Utilizes Convolutional Neural Networks (CNNs) to identify position-invariant patterns within PCAP files, enhancing detection accuracy.
- **Low Computational Overhead**: The model features a compact architecture with few trainable parameters, enabling efficient training and inference using only CPU resources, ideal for edge computing deployments.
- **Continual Learning Architecture**: Maintains consistent, high-quality results by handling data drifts through a continual learning framework.
- **Experience Replay Strategy**: Incorporates experience replay to prevent catastrophic forgetting, ensuring reliable performance over time.
- **Transfer Learning Capabilities**: Demonstrates the ability to transfer learned knowledge to new PCAP flows, enabling seamless expansion to additional data sets.

## 2 Background and Related Work

Traditional methods for analyzing PCAP files have relied heavily on manual inspection and rule-based systems. These techniques are labor-intensive, requiring significant expertise to interpret the complex data structures and are impractical at larger scales due to enormous data volumes. Manual approaches are thorough but time-consuming, while rule-based systems use predefined criteria to detect anomalies and failures in network traffic. However, these systems lack flexibility and struggle to adapt to new network behaviors or patterns, necessitating constant updates as network technologies evolve, which is a maintenance burden. Both methods have scalability issues, struggling to handle the growing volume of data transmitted across networks. This growing volume of data requires automated tools that can efficiently process large datasets and detect faults without significant delays, leading to the adoption of advanced techniques like machine learning models that learn from data and improve over time without the need for constant rule updates.

The application of machine learning (ML) techniques to PCAP data is a relatively new field that has gained traction due to its potential to automate and enhance the accuracy of network analysis. Several studies have employed supervised learning methods, utilizing labeled datasets to train models for classifying network traffic and identifying malware traffic [1]. However, these ML approaches often rely on handcrafted features and labeled data, which can be a significant barrier due to the challenge of maintaining the appropriate features and labeled dataset over time.

Recent advancements in network traffic analysis have introduced innovative methods leveraging Large Language Models (LLMs). These models use advanced natural language processing to interpret network traffic data encoded as text. By treating network traffic and logs as a form of language, LLMs can identify patterns and failures within PCAP files more flexibly and dynamically than traditional rule-based systems [2]. This LLM-based approach offers several advantages, including continuous learning and adaptation to new network behaviors without manual rule updates. It also provides a more nuanced understanding of network data, potentially identifying subtle indications of failure that traditional methods might overlook. Integrating LLMs into PCAP analysis represents a significant advancement, providing a scalable, adaptable, and efficient tool for managing the complexity and volume of network traffic. This study compares our approach (PCAPVision) to LLMCap [2] in terms of prediction speed. Even before the rise of LLM-based approaches, significant progress in network analysis and malicious traffic detection was achieved using computer vision techniques, such as Convolutional Neural Networks (CNNs) or Vision Transformers (ViT).

In 2017, Wei Wang et al. [3] were among the first to apply a representation-based learning approach to malicious traffic classification using CNNs. They divided continuous raw traffic into several discrete units like network flows/sessions, applied traffic cleaning, and converted these units into greyscale images for deep learning models to analyze. Their method created 28x28 pixel images from network flows, which were then fed into a simple CNN, achieving significant accuracy in classifying malware traffic.

Following this, Hussain et al. [4] utilized ResNet, a state-of-the-art CNN model, to detect DoS and DDoS attacks by converting network traffic data into RGB images. This approach achieved remarkable accuracy in identifying various attack patterns, showcasing the potential of CNNs in handling complex network traffic scenarios. In 2022, Agrafiotis et al. [5] advanced the field by incorporating Vision Transformers (ViTs) alongside CNNs to enhance detection accuracy. They used an approach similar to Wei Wang et al., converting traffic data into greyscale images to train both CNNs and ViTs, which yielded better results than standard Random Forest models.

Most recently, Davis et al. [6] proposed a novel approach for malware traffic classification by converting IP header fields from network sessions into 50x50 RGB images. Their method leveraged detailed information within PCAP files to create visual representations processed by deep CNNs. This technique outperformed traditional ML methods across multiple datasets, achieving higher accuracy and demonstrating the robustness of image-based network traffic analysis.

These computer vision-based methods provide a significant improvement over classical feature-based ML approaches, as well as traditional rule-based and manual inspection techniques. They offer a scalable solution for network traffic analysis, capable of processing large volumes of data efficiently and accurately. The ability to automatically learn features from raw traffic data without manual feature extraction represents a major advancement in the field.

While the cited studies primarily focus on malicious traffic detection, our methodology targets failure detection. Other approaches typically preprocess network flows or sessions from the raw traffic data before converting them into images, segmenting data into standardized units for easier analysis. In contrast, our approach bypasses this preprocessing step, allowing direct analysis of raw PCAP files. By avoiding call flow extraction, we achieve significantly faster processing times, to handle large-scale network traffic efficiently and improve real-time network monitoring and failure detection.

As previously mentioned, we use a continual learning framework to maintain our model's robustness over time. Continual learning in neural networks is complex, with various strategies to mitigate issues such as catastrophic forgetting and overfitting. One approach involves using task-specific components, like context-dependent gating, where different model segments are optimized for specific tasks [7]. Alternatively, gradient descent-based methods selectively train certain units, often employing task-oriented attention masks to manage model weight allocation [8]. Another approach uses evolutionary computation, where agents, learned through genetic algorithms, determine parameter subsets for each task [9]. Regularization strategies protect learned parameters from one task during subsequent task training by applying constraints that encourage the model to retain relevant knowledge [10]. The replay method incorporates historical data from previous tasks into the training set to reinforce past learnings [11], although this can be impractical in rapidly changing environments with high throughput requirements due to increased training dataset size. Often, a simple regularization-based approach suffices.

Existing frameworks have shown promise in addressing continual learning challenges. Parisi et al. [12] reviewed various strategies to maintain model performance over time, emphasizing regularization techniques, dynamic architecture approaches, and experience replay methods. In this article, we present a continual learning framework to solve overfitting and catastrophic forgetting.

## 3 Methodology

Our methodology leverages computer vision techniques and a CNN architecture to detect failures in PCAP files. This section details the process of transposing PCAP data into images, the architecture of the CNN used, and the specifics of the training process and dataset employed for failure detection. Additionally, we describe the integration of a continual learning framework that adapts the model dynamically to new data, ensuring sustained performance over time.

### 3.1 Failure Detection

**Convert PCAP files into images**. Inspired by Wei Wang et al. [3], the process to convert PCAP files into images involves several transformation steps. First, the PCAP file is opened in binary mode to read the data as bytes. These bytes are then encoded into a hexadecimal format for readability. Each pair of hexadecimal digits is converted into their decimal equivalents, resulting in an array of decimal numbers corresponding to the original byte values from the PCAP file. This array is reshaped into the desired image size. If the array exceeds the image dimensions, it is truncated; if the array is too small, zeros are appended. Each pixel in the image represents a byte value, ranging from 0 to 255, making it suitable for CNN analysis. In the following, we describe the retained architecture (or more details about the sensitivity analysis, see Appendix A).

**Convolutional Neural Network (CNN) architecture**. The CNN for analyzing PCAP-derived images is structured to handle high-resolution inputs and complex data patterns. Each image is fixed at 1600x1600 pixels, representing bytes from the PCAP file (see Figure 2). **Reshaping and Padding**. The input image is reshaped to 1696x1696 pixels. Reflect padding of 48 pixels on both axes preserves the spatial context of the image and reduces the effects of boundary artifacts during convolution. **Normalization and Dropout**. The pixel values are normalized to a unit interval [0; 1] by dividing by 255, preventing exploding [13] and vanishing gradients [14], thereby improving the stability of the model during training [15]. Following normalization, a two-dimensional dropout with a probability of 0.4 is applied at the outset of the model [16] to enhance model generalization. **Convolutional Layer**. A single convolutional layer with a kernel size of 64x64 and an 8x8 stride optimizes feature extraction. The Exponential Linear Unit (ELU) activation function allows smooth transitions between the activation maps' negative and positive values, capturing intricate PCAP data patterns. To keep the model computationally efficient, 4 kernels are used in the first layer. **Max Pooling**. A two-dimensional max pooling with an 8x8 window reduces the dimensionality and complexity of the feature maps, lowering the number of trainable parameters in subsequent layers and minimizing computational demands. **Flattening and Dense Layers**. Reduced feature maps are flattened into a one-dimensional vector for fully connected processing. The first fully connected (FC) layer includes 256 units and ELU activation integrates learned features. A dropout layer with 0.3 probability reduces overfitting. **Output Layer**. The final layer is a second fully connected layer that outputs through a single unit with a sigmoid activation function. This layer is designed for binary classification, determining the presence or absence of failures in the PCAP file.

**Initial Training Process and Dataset Used**. Our CNN training process for analyzing PCAP files from VoLTE services is structured to ensure robust performance. We use the Adam optimizer [17] with a learning rate of 0.0005. We selected the Binary Cross Entropy [18] loss function as a cost function for the binary classification, identifying PCAP files with at least one failed call-flow (failed PCAP files) or only successful call-flows (successful PCAP files). To prevent overfitting, we adopted a conservative approach for training duration and monitoring, capping at 500 epochs. An early stopping mechanism halts training if no improvement in validation loss is observed for 16 consecutive epochs, preventing overfitting, and ensuring that the model generalizes well on unseen data. The dataset used for training and validation is derived from VoLTE service communications and segmented into training, validation, and testing subsets. The training set comprises 768 successful PCAPs and 454 failed PCAPs, totaling 1,222 files. A failed PCAP contains at least one failed call-flow, a successful PCAP cannot include any failure-classified flows. The validation and test sets both contain 43 successful and 25 failed transmissions, ensuring a balanced mix for model evaluation.

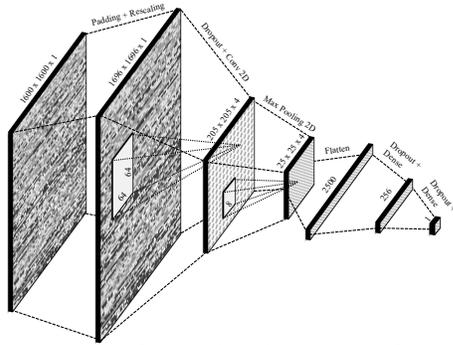

Figure 2: Deep neuronal network architecture used in PCAPVision.

Subsequent training stages introduce continual learning to adapt the model dynamically to new data without forgetting previously learned information. This phase also involves refining and selecting hyperparameters for the final model topology, ensuring the network remains efficient and effective in real-world scenarios.

## 3.2 Continual Learning Framework

Network environments are highly dynamic, with constantly changing traffic patterns and behaviors. This variability can impact model performance, further complicated by vendor-specific rules and constraints. To address these challenges, we employ a continual learning framework that adapts the model weights over time. Due to the large amount of data produced, retaining all PCAP files is impractical, necessitating a strategy for ongoing model fine-tuning without using the entire historical dataset.

Our continual learning framework is composed of several parts, with a cyclical process of monitoring, evaluation, fine-tuning, and updating the PCAPVision model, all within an automated vendor-premises workflow. Figure 3 illustrates our continual learning framework, broken down into four procedures: automatic labeling, data management, daily evaluation and fine-tuning.

We use Call-Flow Analyzer (CFA) to **automatically label new datasets** (Figure 3, P1). CFA processes PCAP files to identify and analyze individual call-flows, detecting failures and diagnosing their causes. This automation replaces the need for manual labeling, significantly increasing the number of annotated PCAP files available for each iteration and ensuring full process automation.

The second procedure corresponds to **Data Management** (Figure 3, P2) collects new PCAP files daily, randomly selecting and labeling 5% for the control dataset. Depending on the volume, additional PCAP files are labeled and added to the daily evaluation set which is then included in the training set for fine-tuning purposes.

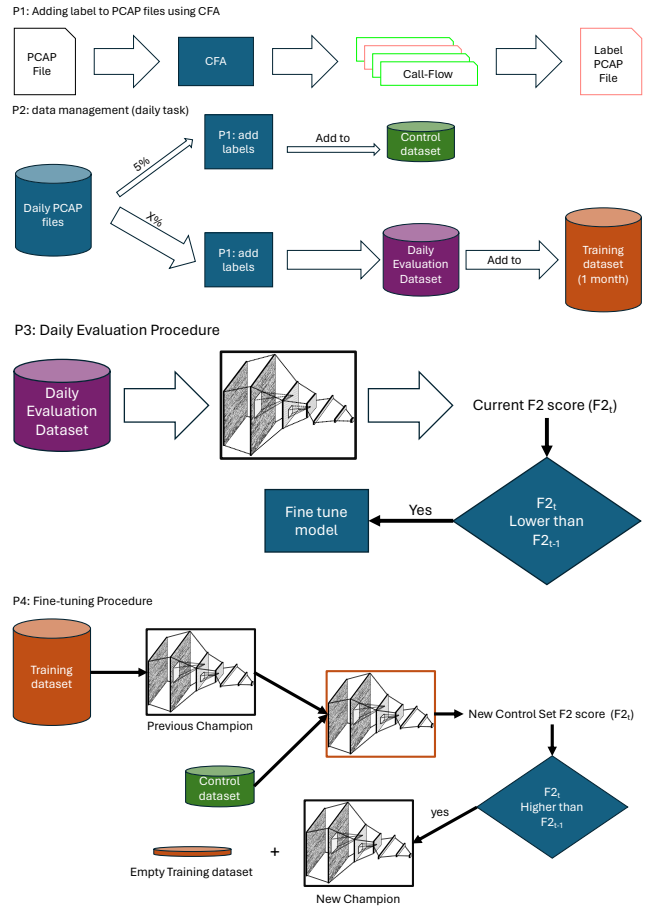

Figure 3: Four procedures (P1-P4) used for continual learning strategy. CFA stands for Call-Flow Analyser.

**Daily Evaluation** evaluates the champion model – our best-performing model to date (see Figure 3, P3) - using metrics (F1-score, F2-score, recall, and precision) with a primary focus on the F2-score. This metric prioritizes minimizing false negatives, ensuring that we capture all failed PCAP files. If the current day's F2-score ($F2_t$) is lower than the previous day's ($F2_{t-1}$), the **fine-tuning** procedure is triggered (P4). All collected metrics are logged to the MLflow [19] tracking server for monitoring and evaluation of the model's performance over time.

When triggered by a drop in the F2-score, the model undergoes fine-tuning with the latest data (see Figure 3, P4). This involves using the accumulated training data collected since the last instance of stable performance. The data collection extends back to the point where no metric decay was observed, ensuring coverage of all relevant periods that might explain the performance decline, such as changes in network behavior. PCAP files are stored for a maximum of one month. During fine-tuning, the most up-to-date model, tagged as "champion" in the MLflow models registry, is selected. This current-best model serves as the baseline for updates. During fine-tuning, we freeze the convolutional layers of the model and focus updates solely on the dense layers, based on the hypothesis that the feature extraction component remains valid for the same task category. We optimize the classification layers to enhance the model's responsiveness to new examples. We reduce the original learning rate by an order of magnitude to prevent large updates during gradient backpropagation, aiming for subtle yet effective model adjustments. Finally, we log the optimal threshold for classifying PCAPs from the newly fine-tuned model. This threshold determines the final class of each PCAP based on its probability score. The result is a refined model, deemed the new champion, ready to be deployed as the current best.

Post fine-tuning, we evaluate the performance of the retrained model candidate using a **control dataset** (Figure 3, P4), a small, curated set (5%) reserved during evaluations. The purpose of this control dataset is to assess the model's susceptibility to catastrophic forgetting to ensure the model retains previously acquired knowledge and to detect if the model is overfitting. Overfitting may occur if the data characteristics used for fine-tuning significantly differ from those of the broader historical data, which could tailor the model parameters to perform well only on the specific, possibly anomalous, features of the fine-tuning dataset and lead to poor generalization on new, unseen data. After evaluating the new model candidate on the control dataset, the new model's performance metrics, particularly the F2-score, are compared against the previous champion. If the new model's F2-score does not exceed the highest recorded score, the model is discarded. Only models demonstrating measurable progress are adopted as the new champion.

We periodically serialize a fraction of historical data into a hold-out set for evaluating the model's ability to retain knowledge without fully retraining. This approach assumes that periodic inference on this set is sufficient to demonstrate the absence of catastrophic forgetting and overfitting. Focusing training on a dataset composed of the most recent PCAP examples allows rapid model updates and suitability for high-throughput production environments. This strategy balances the need for model adaptability with operational demands.

## 4 Results and Discussion

This section presents the findings from our experiments on failure detection in PCAP files using PCAPVision, highlighting the continual learning framework, computational efficiency, and transfer learning capabilities. Finally, we list the limitations of PCAPVision. In this section, we only consider the best architecture found. For a complete overview of tested architectures, see Deep Learning Architectures in Appendix A. We conducted a sensitivity analysis to evaluate the performance of different types of architectures including Recurrent Neural Networks (RNN) which provided poor results (F2-score: 0.48) compared to the CNN model (F2-score: 0.89) on the VoLTE dataset.

### 4.1 Evaluation of the Continual Learning Framework in a Lab Environment

We implemented the continual learning framework in a laboratory setting, where daily VoLTE tests generate 40 to 1,000 PCAP files per day. Each PCAP file is retained for a period of one month before becoming inaccessible. This environment allowed us to evaluate our continual learning framework over a period slightly exceeding one month to determine if the framework could autonomously improve its performance over time without any external intervention. The initial model, V1, exhibited moderate performance with F2 scores between 0.6 and 0.8, trained on our internal dataset without fine-tuning. On the fourth day, a drop in F2 score triggered fine-tuning, making V2 the new champion as it had no prior control dataset scores. The next day, another performance drop triggered a fine-tuning, resulting in model V3 becoming the new champion. However, subsequent models failed to surpass V3's performance on the control dataset until March 23, when V4 was established as the new champion. A notable period of instability occurred between April 1 and April 6, characterized by two significant drops in performance, primarily due to the low number of tests (approximately 50 PCAP files) compared to other days (around 300 PCAP files). The tests conducted during this period were substantially different from those conducted previously. Attempts to fine-tune the model during this time did not improve control set performance, and no new champions were selected. Later in April, models V5 and V6 were appointed as new champions, both demonstrating high stability and enhanced performance, with F2 scores exceeding 0.8 on both the control and evaluation datasets. This outcome demonstrates the effectiveness of our continual learning framework in achieving convergence towards high performance and stability over time.

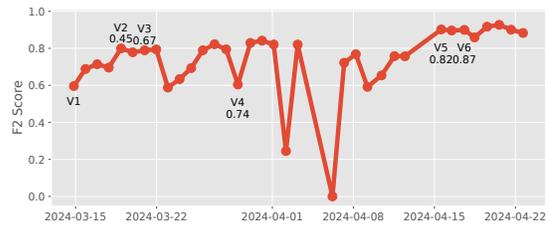

Figure 4: F2-Score obtained using the daily evaluation procedure on an isolated lab environment. The mention *VX* denotes when a new version has been used as a new champion. The number below the version indicates the F2-Score on the control dataset.

### 4.2 Computational Efficiency Analysis of PCAPVision Compared to Other Methods

To illustrate the low computational overhead, we compared processing times across four systems: PCAPVision, LLM-Cap [2], CFA, and Subject Matter Experts (SMEs). The task was to detect call-flow failures in PCAP files, excluding the identification of specific failure reasons. While PCAPVision focuses solely on detecting failures, LLMCap and CFA offer broader functionalities that could complement PCAPVision for detailed PCAP file analyses post-failure detection.

We used VoLTE PCAP files of various file sizes (mean size = 1.03 Mb, std = 0.7) and reported average processing times to reflect general computational effort. To compare speed, LLMCap and PCAPVision tests were run on an AWS EC2 *g5.4xlarge* instance. CFA uses a dedicated Kubernetes cluster of 8 cores. We ensured that each system was tested under its optimal setup to provide a fair comparison of computational time.

For PCAPVision, the inference process involves converting the binary PCAP file into an image and then using the model for inference. Our analysis revealed that preprocessing (PCAP to image conversion) accounted for 98% of the total computational time, with the inference process consuming just 2%. This highlights the significant computational demand of the preprocessing step in the PCAPVision approach. Table 1 shows that PCAPVision detects failures significantly faster than other automated solutions, achieving speeds one to two orders of magnitude greater. Compared to manual analysis, PCAPVision's speed advantage extends to several orders of magnitude.

| Method | Inference Duration |
| --- | --- |
| PCAPVision | 0.45 [s] |
| LLMCap [2] | 9.5 [s] |
| CFA | 10-20 [s] |
| SME | +/- Minutes |

Table 1: Computational inference time in seconds, and resources for each method.

In high-volume production environments, PCAPVision can initially filter PCAP files to identify potential failures. Once a failure is detected, more detailed analyses can be conducted using LLMCap or CFA. This staged approach allows efficient initial screening by multiple instances of PCAPVision at the network edge, followed by focused, deeper analysis only on flagged PCAP files. This optimizes resource use and ensures intensive processing is only applied to relevant data.

### 4.3 Transfer Learning

In this study, we explored the transfer learning capabilities of a pretrained model, initially trained on the VoLTE service dataset. We extended the evaluation to the Mobility Management dataset to assess its adaptability and generalization capabilities. The Mobility Management dataset includes 674 successful and 230 failure PCAPs for training, 85 successful and 34 failure PCAPs for validation, and 85 successful and 38 failure PCAPs for testing. We conducted several experiments to assess the model's performance on the Mobility Management dataset (Figure 5). First, as a reference, we evaluated the strategy of assigning the label "failure" based on a prior probability of 0.31 based on the test set, yielding an F2-score of 0.18. Next, we applied the pretrained model directly to the new dataset with no fine-tuning, achieving an F2-score of 0.42. Compared to the random strategy, the zero-shot performance indicates that the model had learned useful information from the VoLTE dataset, which was partially transferable to the Mobility Management dataset. We then fine-tuned only the dense layers (prediction part of the network) of the pretrained model, resulting in an F2-Score of 0.56. This approach improved performance compared to zero-shot but was not as effective as training the entire model from scratch, which achieved an F2-Score of 0.67. This suggests that the feature maps learned from the VoLTE dataset did not generalize well to the new dataset, necessitating retraining. The most significant improvement was observed when we fine-tuned the entire pretrained model on the Mobility Management dataset, resulting in an F2-score of 0.72. This performance surpassed that of training the full model from scratch, indicating a positive transfer of knowledge between the two datasets. Fine-tuning the entire model allowed us to leverage the pre-existing knowledge from the VoLTE dataset while effectively adapting to the new data context.

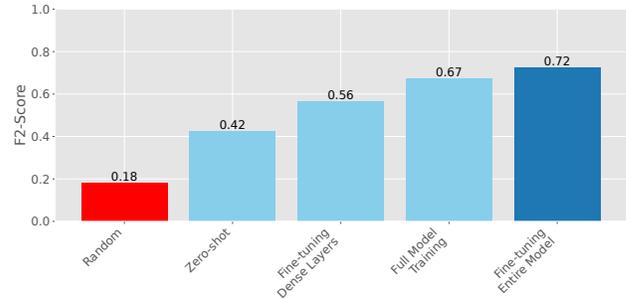

Figure 5: Comparison of Obtained F2-Score using different transfer learning strategies.

### 4.4 Limitations of PCAPVision

PCAPVision faces two main limitations in its current implementation. Firstly, its accuracy is closely linked to the accuracy of CFA, which automatically labels PCAP files in both training and control datasets. Inaccuracies in CFA's labeling directly affect PCAPVision's performance. Secondly, PCAPVision is currently limited to processing files up to 2.5 MB, with excess data truncated. To address this, adjustments to the input size of the first layer or segmentation of larger files into smaller chunks using tools like Wireshark's editcap [20], can ensure complete analysis without data loss.

## 5 Conclusions

This work introduces PCAPVision, an innovative approach for detecting failures in PCAP files using computer vision and Convolutional Neural Networks (CNNs). By converting PCAP data into images, it harnesses the pattern recognition capabilities of CNNs for efficient and accurate network traffic analysis, surpassing traditional parsing methods in speed and precision. A key innovation of PCAPVision is its continual learning framework, which maintains robust performance over time by dynamically adapting to new data. Using techniques like experience replay and adaptive regularization, the framework mitigates catastrophic forgetting, ensuring the model remains effective in evolving network environments. Demonstrating its versatility, PCAPVision was initially trained on a VoLTE dataset and subsequently fine-tuned on a Mobility Management dataset, highlighting its transfer learning capabilities across various network services.

In high-rate, high-volume production environments, PCAPVision can significantly reduce storage costs by filtering out non-failure PCAPs, ensuring that only relevant data are stored and analyzed. This efficiency optimizes resource usage and accelerates the failure detection process, facilitating rapid identification and remediation of network issues.

Future work could explore ensemble weighted methods to improve PCAPVision's accuracy and reliability. By integrating multiple models specialized in different aspects of network traffic analysis, overall performance could be markedly improved. Expanding the application of PCAPVision to other domains of network traffic analysis could broaden its utility and impact.

## Acknowledgments
The authors gratefully acknowledge Dina Bennett for their invaluable assistance and insightful feedback during the preparation of this manuscript.

## A  Deep Learning Architectures

We evaluated various architectures to detect failure in PCAP files, aiming to identify the optimal, minimal model structure. This process involved evaluating different configurations, including the number and types of layers, to identify the most effective topology for our specific tasks. Our training dataset consists of 1,222 PCAP files containing VoLTE call-flows (see Initial Training Process and Dataset Used subsection 3.1).

| Model Architecture | Test Recall | Test Precision | Test F2 |
|---|---|---|---|
| 1-layer LSTM | 0.32 | 0.60 | 0.36 |
| 3-layer LSTM | 0.45 | 0.63 | 0.48 |
| 1-layer Conv2D & BatchNorm | 0.79 | 0.34 | 0.63 |
| 1-layer Conv2D & BatchNorm & MaxPool | 0.64 | 0.77 | 0.66 |
| 1-layer Conv2D & MaxPool & Dropout | 0.84 | 0.82 | 0.84 |
| WrapPad & 1-layer Conv2D & MaxPool & Dropout | 0.86 | 0.77 | 0.84 |
| WrapPad & Dropout & 1-layer Conv2D & MaxPool & Dropout | 0.88 | 0.92 | 0.89 |

Table 2: Comparison of different model architectures based on Test Recall, Precision, and F2 score.

In this supplementary section, we outline the experiments incorporated in the sensitivity analysis. Initially, we explored a Long Short-Term Memory (LSTM) model [21], where each PCAP file is transformed into a one-dimensional signal. The length of the PCAP file dictates the number of time-steps in the LSTM, providing flexibility in handling varying PCAP file sizes due to the model's invariance to time-step count. However, training results indicated unsatisfactory convergence, suggesting that the diverse sizes of PCAP files might hinder optimization and contribute to training instability. Additionally, we observed issues such as vanishing or exploding gradients [22], possibly triggered by anomalous byte sequences inherent in the PCAP samples. Next, we explored converting PCAP samples into 2-dimensional images, employing a convolution-based model architecture. We hypothesized that both the temporal information and contextual details in PCAP data could enhance failure prediction capabilities, supported by the observation that consecutive groups of packets provide insights into the timing and sequence of network events, which are crucial for analyzing network performance metrics and addressing issues related to latency, packet loss, and jitter. Our initial model configuration included a single 2-dimensional convolutional layer, serving as a feature extractor, followed by a Flatten layer and two fully connected (FC) layers for binary classification. During training, we conducted a quantitative analysis to assess model generalization and the extent of overfitting using a hold-out validation set. Results indicated overfitting on the training data, prompting us to introduce max pooling and a dropout operation prior to the FC layers to enhance model generalization. We experimented with adding mirror padding to the PCAP images to provide additional contextual information for the convolutional kernels, addressing the challenge of edge pixels in a row-wise converted image, which are often correlated with their opposite counterparts. Adding *mirror* padding helped mitigate the loss of potentially valuable information across image boundaries. The inclusion of padding improved contextual awareness, increased the speed of propagation, and allowed for more efficient batching. We also placed a dropout layer between the padding and convolution operations to further reduce overfitting, enhancing the performance of our model.

We employ the Adam optimizer [17] with a learning rate of 0.0005 across all models. The loss function used is Binary Cross Entropy [18], with the decision threshold determined by analyzing the precision-recall trade-off on the validation set. Given the high impact of false negatives in failure prediction tasks, we emphasize recall in the threshold selection process and adopt the F2-score to prioritize this metric. Table 2 summarizes our sensitivity analysis and grid search results for determining the optimal model architecture. The final architecture of our model (described in the section Methodology 3) has two primary components: the feature extractor and the predictor. The feature extractor includes mirror padding, a dropout layer, a single convolutional layer with four kernels, followed by max pooling and flattening operations. The predictor comprises two blocks, each containing a dropout layer and a FC layer.